\documentclass[aip,jap,reprint,amsmath,amssymb,floatfix]{revtex4-2}

\usepackage{graphicx}       % figures
\usepackage{dcolumn}        % align table columns on decimal point
\usepackage{bm}             % bold math
\usepackage{amsmath}
\usepackage{amssymb}
\usepackage{siunitx}        % SI units
\usepackage{physics}        % bra-ket notation
\usepackage{xcolor}

\usepackage{hyperref}
\hypersetup{colorlinks=true,linkcolor=blue,citecolor=blue,urlcolor=blue}

\newcommand{\kp}{$\mathbf{k}\cdot\mathbf{p}$}
\newcommand{\LL}{Localization Landscape}
\newcommand{\WW}{Wigner--Weyl}
\newcommand{\AlGaN}{Al$_{x}$Ga$_{1-x}$N}

\begin{document}

\preprint{AIP/123-QED}

\title{Efficient Analysis of Carrier Transport and TM-TE Emission in AlGaN UVC LEDs via Multi-band Localization Landscape Theory}

\author{Yu-Ming Chang}
\affiliation{Graduate Institute of Photonics and Optoelectronics and Department of Electrical Engineering, National Taiwan University, Taipei 10617, Taiwan}

\author{Ping-Jie Zhuang}
\affiliation{Graduate Institute of Photonics and Optoelectronics and Department of Electrical Engineering, National Taiwan University, Taipei 10617, Taiwan}

\author{Marcel Filoche}
\affiliation{Institut Langevin, ESPCI Paris, PSL University, CNRS, 75005 Paris, France}

\author{Claude Weisbuch}
\affiliation{Laboratoire de Physique de la Mati\`ere Condens\'ee, CNRS, \'Ecole polytechnique, Institut Polytechnique de Paris, 91120 Palaiseau, France}
\affiliation{Materials Department, University of California, Santa Barbara, California 93106, USA}

\author{James S. Speck}
\affiliation{Materials Department, University of California, Santa Barbara, California 93106, USA}

\author{Yuh-Renn Wu}%
\email{yrwu@ntu.edu.tw}
\affiliation{Graduate Institute of Photonics and Optoelectronics and Department of Electrical Engineering, National Taiwan University, Taipei 10617, Taiwan}

\date{\today}

% ------------------------------------------------------------------
\begin{abstract}
AlGaN-based UVC LEDs (220--250~nm) suffer from poor hole confinement and strain-induced $\ket{Z}$-band dominance at high Al content ($>$60\%), leading to increased TM emission and reduced external quantum efficiency (EQE). While conventional \kp{} models combined with Schr\"{o}dinger, Poisson, and drift-diffusion solvers are widely used to study optical transitions, they are computationally expensive. In this work, we apply the multi-band \LL{} (LL) model, including the effect of strain, as an alternative that replaces the eigenvalue problem to efficiently capture quantum effects and carrier localization. Using the 3D multi-band LL model with the \WW{} formalism, we reproduce emission and absorption spectra trends similar to the results in 3D \kp{} calculations, but with significantly reduced simulation time. The polarization ratio also agrees well with published experimental results across a wide spectral range. Furthermore, we analyze electrical characteristics such as band structure, polarization switching, and carrier confinement under alloy fluctuations and strain. This multi-band LL-based approach provides a fast and reliable solution for understanding and optimizing UVC LED performance.
\end{abstract}

\maketitle

% ==================================================================
\section{\label{sec:intro}Introduction}
% ==================================================================

Aluminum gallium nitride (AlGaN)-based deep ultraviolet (UVC) light-emitting diodes (LEDs), emitting in the $220$--$300$~nm wavelength
range, are promising candidates for next-generation germicidal light sources~\cite{amano2020,Zollner_2021}. Their compact size, absence of hazardous
mercury~\cite{khan2008,kneissl2019}, and emission spectrum overlapping with the DNA absorption peak~\cite{nagasawa2018,song2018}
($250$--$265$~nm) make them particularly suitable for disinfection and biomedical applications. Furthermore, emission around 220--230~nm
features a smaller skin depth to human cells, making it less harmful to the human body.

\begin{figure}[t]
  \includegraphics[width=\columnwidth]{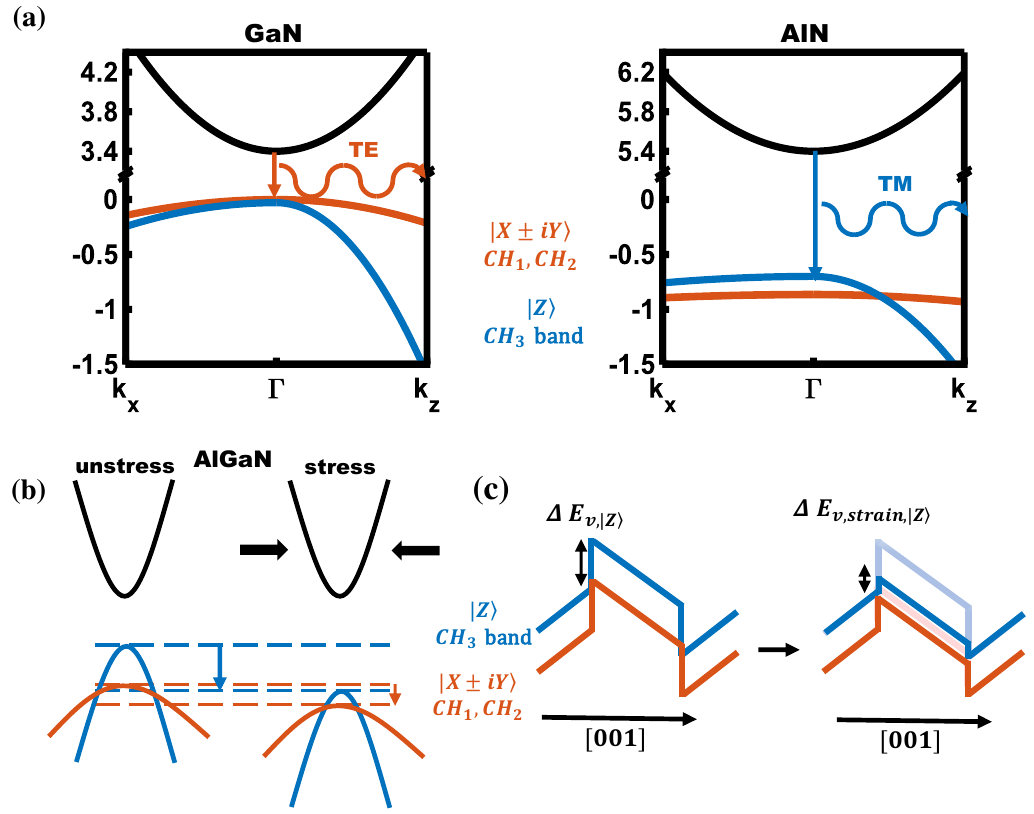}
  \caption{\label{fig:banddg}(a) The band structure in GaN and AlN showing polarization mode transitions from TE- to TM-dominant emission as the valence band ground state changes from $\ket{X\pm Y}$ to $\ket{Z}$. (b) Schematic valence-band diagram of AlGaN with and without in-plane compressive stress. (c) When the compressive stress  (deformation-potential term) is included in the QW, the valence-band edges (CH$_1$--CH$_3$) in the QW shift downward in energy, effectively making QW shallower as shown in (c).}
\end{figure}

However, achieving efficient UVC emission from AlGaN structures remains a challenge. High aluminum composition is required for deep UV operation, which exacerbates several physical limitations. For example, the high acceptor ionization energy in p-type AlGaN leads to low hole concentrations~\cite{amano2020}. As the aluminum composition increases, both electron and hole transport across the active region become more difficult. Because electrons generally have higher mobility than holes~\cite{hopfner2024}, electrons commonly flow through the quantum wells before holes arrive at the radiative-recombination region, leading to reduced recombination efficiency. Furthermore, AlGaN-based multiple quantum well (MQW) LEDs exhibit alloy fluctuations~\cite{shen2021, odonovan2024, shen2022, schulz2015, weisbuch2021} in both quantum wells and quantum barriers, which seriously affect quantum confinement effects and transport. The alloy fluctuations in the quantum barriers create percolation paths for electrons, which may further worsen the electron overflow due to poor hole injection~\cite{su2017,yang2012}. In addition, with higher Al composition, the $\ket{Z}$ band dominates as the ground state valence band, which has a lighter effective mass along the $z$-direction as shown in Fig.~\ref{fig:banddg}(a), and makes the transport of holes in the growth direction easier. However, the $\ket{Z}$ band (CH$_3$) also leads to strong TM polarization, which is detrimental for light extraction. Hence, it is relatively complicated to understand the device physics in AlGaN-based UVC-LEDs.

To suppress the TM polarization, it is known that when under compressive stress, the $\ket{Z}$ valence band will moves down as shown in Fig.~\ref{fig:banddg}(b), which reduces the TM polarized light emission. To achieve this during the fabrication process, AlN substrates, or templates with an Al content higher than that of the well, are often employed to induce biaxial compressive stress~\cite{suzuki1996} in the AlGaN QWs. Under the biaxial compressive stress, the $\ket{X\pm iY}$ bands are more likely to be the ground state band, especially when the Al composition in the AlGaN QW is smaller than 60\%. Furthermore, the compressive stress introduces additional piezoelectric polarization. Combined with the intrinsic spontaneous polarization, these effects lead to strong internal polarization-related electric fields within the quantum well. In c-plane GaN/AlGaN quantum wells, strong polarization-induced internal electric fields give rise to a pronounced quantum-confined Stark effect (QCSE), evidenced by a redshift of the optical transition and field-dependent excitonic absorption/photoluminescence~\cite{leroux1998,deguchi1999}. These built-in fields tilt the band profile across the well, spatially separating electrons and holes, reducing their wave-function overlap, and further reduces the radiative recombination rate.

As mentioned, for higher Al composition, the dominant state is the $\ket{Z}$ state as shown in Fig.~\ref{fig:banddg}(a). However, if compressive stress is introduced in the AlGaN QW during epitaxial growth, it may shift the relative position of the $\ket{Z}$ state as shown in Fig.~\ref{fig:banddg}(b). Under high compressive strain, the $\ket{Z}$ state moves down further, and it is possible to switch from TM to TE emission if sufficiently strong compressive stress is applied. The Al composition at which TM and TE polarization modes switch differs according to different buffer-layer lattice sizes~\cite{banal2009,nam2004}. If the buffer layer is thick enough that stress is fully relaxed from the substrate, then the stress of the AlGaN QW/QB layer is highly related to the buffer layer's lattice size after stress relaxation. Different choices of substrate or buffer layer will strongly affect the band movement the $\ket{Z}$ and the $\ket{X\pm iY}$ state, and further affect the optical polarization of emitted light. Studies~\cite{ryu2013, northrup2012} have shown that TM-polarized light propagates parallel to the quantum well plane, making light extraction more challenging along the c-axis direction.

In multiple quantum-well (MQW) structures, the interplay between piezoelectric polarization and strain-induced band changes becomes increasingly important~\cite{liu2025,xu2019}. If simulations consider only the piezoelectric polarization and neglect the deformation-potential effect, one may obtain inaccurate results. To illustrate this, Fig.~\ref{fig:banddg}(c) shows valence band structures with and without inclusion of the deformation-potential term. Neglecting the deformation potential, the band profile exhibits higher potential barriers for holes, leading to stronger confinement and inefficient inter-well transport. In contrast, including the deformation term introduces a strain-induced shift in the band structure, forming a shallower quantum well that makes holes less confined in the QW . It then enhances vertical QW-to-QW transport, where the hole might be easier to escape and move to the next QW. Hence, it is essential to include both polarization fields and strain-related band-structure modifications in simulation models to realistically describe carrier dynamics and optimize UVC LED performance.

In ternary III-nitride quantum wells (e.g., $\text{Al}_{x}\text{Ga}_{1-x}\text{N}$ or $\text{In}_{x}\text{Ga}_{1-x}\text{N}$), random alloy fluctuations are intrinsic because cation sites are occupied discretely and statistically rather than forming a perfectly uniform virtual crystal. This inevitably produces three-dimensional (3D) spatial fluctuations in band edges (local band gap and band offsets), and---through composition-dependent lattice mismatch and piezoelectric coupling---induces local variations of strain and polarization fields, yielding a rugged 3D potential landscape that cannot be represented by a purely 1D averaged quantum well. Atomistic calculations have repeatedly shown that such alloy-induced disorder drives strong hole localization, leading to inhomogeneous broadening and modified optical transition strengths compared with virtual-crystal treatment~\cite{schulz2015,finn2022}. For AlGaN quantum wells in UV LEDs, random alloy fluctuations alone can produce strong carrier localization and, very importantly, can shift the TE-TM polarization crossover relative to virtual-crystal predictions, directly impacting light-extraction considerations for deep-UV LEDs~\cite{shen2022,finn2022}. The 3D nature of alloy fluctuations enables (i) lateral percolative paths and (ii) effective barrier lowering/tunneling-assisted transport that alter carrier injection, inter-well transport, and leakage relative to smooth laterally invariant band profiles.

\begin{figure}[t]
  \includegraphics[width=0.8\columnwidth]{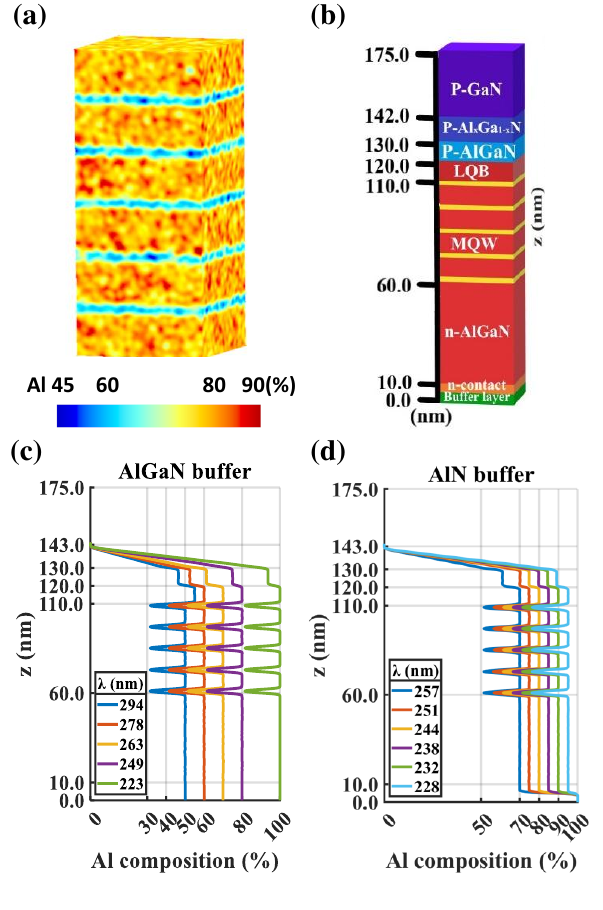}
  \caption{\label{fig:structure}(a) Three-dimensional aluminum composition distribution in AlGaN quantum wells based on the random alloy fluctuations. (b) The simulated structure shown is the MQW group with 2~nm AlGaN QW/10~nm AlGaN QB, 10~nm p-AlGaN and top 12~nm AlGaN graded-composition layer doped layer to help the hole carrier injection, and 33~nm p-GaN to form the ohmic contact with metal contact. (c) Vertical aluminum-composition profiles along the growth direction used with an AlGaN buffer for the five LED structures, targeting emission at 294, 278, 263, 249, and 223~nm. (d) Vertical aluminum-composition profiles used with an AlN buffer for the six LED structures, targeting emission at 257, 251, 244, 238, 232, and 228~nm. In our simulations, the Al composition in QW/QB always differs by 20\%. The detailed Al composition profile for each LED structure is provided in the Appendix.}
\end{figure}

In this context, the \LL{} (LL) theory~\cite{filoche2017a,li2017,piccardo2017,arnold2016} provides an efficient way to simulate disordered systems. It can efficiently realize the quantum potential and quantum-corrected carrier statistics within a drift-diffusion framework, enabling device-scale simulations that retain key disorder-induced quantum effects without solving full Schr\"{o}dinger eigenvalue problems for every bias point. As mentioned earlier, the relative position of $\ket{X\pm Y}$ and $\ket{Z}$ bands strongly depends on both the Al composition and local strain state. When random alloy fluctuation is considered, the strain and composition at each local site vary as shown in Fig.~\ref{fig:structure}(a), where the dominant ground state might switch between $\ket{X\pm Y}$ and $\ket{Z}$ locally due to composition variation. Theoretically, to handle the strain-induced deformation potential shift, the \kp{} model is typically used. However, solving the eigenvalue problem for the 3D \kp{} problem within a large simulation domains is time-consuming. In this paper, to address this problem, we propose the multi-level Localized landscape model to solve the $\ket{X\pm Y}$ and $\ket{Z}$ bands separately, including the influence of strain and composition variation, which enables us to estimate their impact on carrier transport in the UVC device system.

Following Fig.~\ref{fig:structure}(b), we analyze the entire LED stack---including the active and the contact regions---using a hybrid framework. We adopt a simulation framework based on the LL model, which enables quantum effects to be incorporated into the classical drift-diffusion formalism by replacing the eigenvalue problem with an effective quantum potential. The LL method, enhanced with the \WW{} formalism~\cite{banon2022}, allows the estimation of carrier localization, emission spectra, and polarization characteristics without explicitly solving the \kp{} Schr\"{o}dinger eigenvalue problem. To account for the influence of strain, we apply a multi-band \LL{} model where different bands are solved separately with the energy shift induced by strain. This allows us to further analyze the TE/TM emission status without solving the $6\times 6$ \kp{} method. The band-edge potential and carrier density used in this calculation are obtained from a 3D drift-diffusion-current-continuity (3D-DDCC) solver~\cite{shen2022,filoche2017a,li2017,piccardo2017}, which solves the Poisson and drift-diffusion equations on a finite-element mesh(FEM), accounting for material composition, doping, and electrostatics across the full LED structure. The local strain was compute with a 3D FEM-based strain-stress solver, where details are described below.

To compare the accuracy of the modified multi-band localized landscape model, the traditional 3D $6\times 6$ valence-band \kp{} method including random alloy effects~\cite{wu2009,shen2022} was also compared with this modified multi-band LL solver. However, due to the intensive computing resources needed in the 3D \kp{} method for the eigenvalue problem, we could only solve the active QW region in the \kp{} solver after the solver has converged, which only aims to calculate the emission spectrum. It is nearly impossible to solve the self-consistent 3D \kp{} method and Poisson equation simultaneously due to demanding computational resources required(see below).

This combined modeling framework allows investigation of how random alloy fluctuations, strain, and quantum barrier design impact the polarization characteristics and carrier injection efficiency of AlGaN UVC LEDs for the 220-300~nm emission. The method provides an efficient and physically accurate tool for understanding internal quantum efficiency(IQE) variation with emission wavelength and Al composition.

% ==================================================================
\section{\label{sec:method}Methodology}
% ==================================================================

\begin{figure}[t]
  \includegraphics[width=0.8\columnwidth]{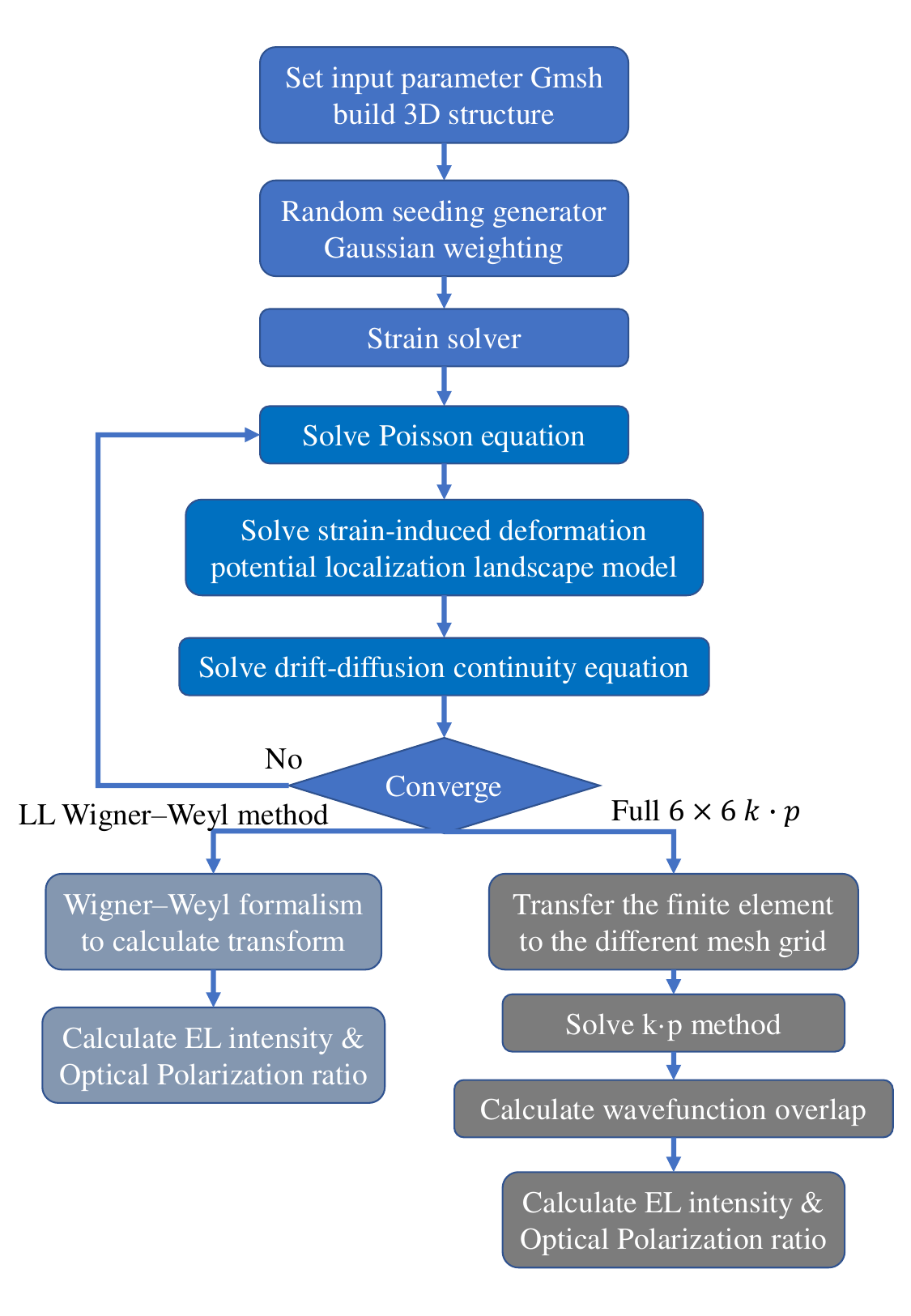}
  \caption{\label{fig:flowchart}Simulation flowchart of the 3D-DDCC program.}
\end{figure}

A flowchart of the simulations is shown in Fig.~\ref{fig:flowchart}. The device geometry and finite-element mesh were generated using Gmsh~\cite{geuzaine2009}, where all layers of the LED is explicitly represented. Material and transport parameters used in the device simulation are summarized in Table~\ref{tab:transport}.

In this work, we employ a three-dimensional drift-diffusion charge-control (3D-DDCC) framework as the core device simulator for AlGaN-based UVC LEDs with random alloy fluctuations~\cite{chen2018}.

To capture microscopic disorder effects in ternary AlGaN, random alloy fluctuations are explicitly included throughout the device (except the binary GaN contact layers). In each mesh element, the Al and Ga compositions were assigned by drawing random seeds on an underlying atomic grid and performing a Gaussian-weighted average over nearby sites. This procedure yields smooth yet random spatial profiles that mimics experimentally observed naturally alloy disorder, rather than ideal, perfectly uniformly averaged compositions. The resulting composition map is interpolated onto the finite-element mesh and mapped to local band-edge energies, effective masses, dielectric constant, polarization coefficients, and elastic stiffness constants. As a consequence, both the conduction and valence band profiles, as well as the polarization charge density $\rho_\text{pol}$, inherit the microscopic alloy fluctuations and directly influence carrier transport and radiative recombination.

In AlGaN-based heterostructures, lattice mismatch induces elastic strain that modifies band edges and generates piezoelectric charges. The elastic strain-stress solver on the same 3D mesh provides the nodal displacement field from which $\varepsilon_{ij}$ is obtained by finite-element post-processing for coupling to polarization and deformation-potential shifts.

\subsection{\label{sec:strain}Strain and Piezoelectric Polarization}

The relation between stress and strain is written in Voigt notation as
\begin{equation}
  \label{eq:voigt}
  \begin{pmatrix}
    \sigma_{xx} \\ \sigma_{yy} \\ \sigma_{zz} \\
    \sigma_{yz} \\ \sigma_{zx} \\ \sigma_{xy}
  \end{pmatrix}
  =
  \begin{pmatrix}
    c_{11} & c_{12} & c_{13} & 0 & 0 & 0 \\
    c_{12} & c_{11} & c_{13} & 0 & 0 & 0 \\
    c_{13} & c_{13} & c_{33} & 0 & 0 & 0 \\
    0 & 0 & 0 & c_{44} & 0 & 0 \\
    0 & 0 & 0 & 0 & c_{44} & 0 \\
    0 & 0 & 0 & 0 & 0 & \frac{c_{11}-c_{12}}{2}
  \end{pmatrix}
  \begin{pmatrix}
    \varepsilon_{xx} \\ \varepsilon_{yy} \\ \varepsilon_{zz} \\
    \varepsilon_{yz} \\ \varepsilon_{zx} \\ \varepsilon_{xy}
  \end{pmatrix},
\end{equation}
where $c_{ij}$ are the stiffness coefficients of wurtzite III-nitrides, interpolated between GaN and AlN according to the local Al composition.

The piezoelectric polarization is then computed from the strain tensor $\varepsilon_{ij}$ using the wurtzite piezoelectric tensor,
\begin{align}
  \label{eq:piezo}
  P_\text{piezo} &=
  \begin{pmatrix}
    0 & 0 & 0 & 0 & e_{15} & 0 \\
    0 & 0 & 0 & e_{15} & 0 & 0 \\
    e_{31} & e_{31} & e_{33} & 0 & 0 & 0
  \end{pmatrix}
  \begin{pmatrix}
    \varepsilon_{xx} \\ \varepsilon_{yy} \\ \varepsilon_{zz} \\
    \varepsilon_{yz} \\ \varepsilon_{zx} \\ \varepsilon_{xy}
  \end{pmatrix}
  \nonumber \\
  &=
  \begin{pmatrix}
    e_{15}\,\varepsilon_{zx} \\
    e_{15}\,\varepsilon_{yz} \\
    e_{31}(\varepsilon_{xx}+\varepsilon_{yy}) + e_{33}\,\varepsilon_{zz}
  \end{pmatrix},
\end{align}
and combined with the spontaneous polarization to yield the total polarization field.

\begin{table}[t]
\caption{\label{tab:mat}Material parameters for GaN and AlN used in the simulation~\cite{romanov2006}.}
\begin{ruledtabular}
\begin{tabular}{lcc}
  \textbf{Parameter} & \textbf{GaN} & \textbf{AlN} \\
  \hline
  $a$ (\AA) & 3.189 & 3.112 \\
  $c$ (\AA) & 5.185 & 4.982 \\
  $c_{11}$ (GPa) & 36.7 & 39.6 \\
  $c_{12}$ (GPa) & 13.5 & 13.7 \\
  $c_{13}$ (GPa) & 10.3 & 10.8 \\
  $c_{33}$ (GPa) & 40.5 & 37.3 \\
  $c_{44}$ (GPa) & 9.5  & 11.6 \\
  $c_{66}$ (GPa) & 11.6 & 12.95 \\
  $e_{31}$ (C/m$^{2}$) & $-0.49$ & $-0.58$ \\
  $e_{33}$ (C/m$^{2}$) & $0.73$  & $1.55$ \\
  $e_{15}$ (C/m$^{2}$) & $-0.40$ & $-0.48$ \\
\end{tabular}
\end{ruledtabular}
\end{table}

\subsection{\label{sec:kp}Six-Band $k\cdot p$ Model}

For accurately analyzing valence band structure and optical light polarization properties in III-nitride quantum wells, a three-dimensional $6\times 6$ \kp{} model~\cite{chuang1996,suzuki1995,wu2009} is needed, which includes the effects of spin-orbit interaction, crystal field splitting, and strain-induced deformation potentials. This model enables the evaluation of band mixing, hole wavefunctions, and polarization characteristics under various strain and composition profiles. As mentioned earlier, this model is time-consuming, and we propose the simplified multi-band LL model to replace \kp{} for efficient computation. Nevertheless, in this paper the $6\times 6$ \kp{} model is also applied and solved for eigenvalues for comparison of the accuracy of the LL model.

\begin{widetext}
The valence band Hamiltonian is constructed in the following block matrix form~\cite{ghosh2002}:
\begin{equation}
  \label{eq:Hv}
  H^{v} =
  \begin{pmatrix}
    F+E_v & 0 & -H^* & 0 & K^* & 0 \\
    0 & G+E_v & \Delta & -H^* & 0 & K^* \\
    -H & \Delta & \lambda+E_v & 0 & I^* & 0 \\
    0 & -H & 0 & \lambda+E_v & \Delta & I^* \\
    K & 0 & I & \Delta & G+E_v & 0 \\
    0 & K & 0 & I & 0 & F+E_v
  \end{pmatrix}
  \begin{pmatrix}
    \Phi_1^v \\ \Phi_2^v \\ \Phi_3^v \\
    \Phi_4^v \\ \Phi_5^v \\ \Phi_6^v
  \end{pmatrix},
\end{equation}
\end{widetext}
where each parameter is defined as:
\begin{align}
  F &= \Delta_1 + \Delta_2 + \lambda + \theta, \\
  G &= \Delta_1 - \Delta_2 + \lambda + \theta, \\
  H &= i\!\left(A_6 k_z k_+ + A_7 k_+ + D_6 \varepsilon_{z+}\right), \\
  I &= i\!\left(A_6 k_z k_+ - A_7 k_+ + D_6 \varepsilon_{z+}\right), \\
  K &= A_5 k_+^2 + D_5 \varepsilon_+, \quad \Delta = \sqrt{2}\,\Delta_3, \\
  \lambda &= A_1 k_z^2 + A_2 k_\perp^2 + D_1\varepsilon_{zz}
            + D_2(\varepsilon_{xx}+\varepsilon_{yy}), \\
  \theta  &= A_3 k_z^2 + A_4 k_\perp^2 + D_3\varepsilon_{zz}
            + D_4(\varepsilon_{xx}+\varepsilon_{yy}), \\
  \varepsilon_+ &= \varepsilon_{xx} - \varepsilon_{yy} + 2i\varepsilon_{xy}, \\
  \varepsilon_{z+} &= \varepsilon_{xz} + i\varepsilon_{yz}, \\
  k_\perp^2 &= k_x^2 + k_y^2, \quad k_+ = k_x + ik_y.
  \label{eq:kp_params}
\end{align}

The strain tensor $\varepsilon$ and valence band edge $E_v$ are obtained from the 3D-DDCC strain and Poisson solver. The parameters $D_1$ to $D_6$ are deformation potential coefficients, and $A_1$ to $A_7$ are equivalent to the valence-band parameters and determine the hole effective masses. $\varepsilon_{lm}$ and $k_l$ ($l,m = x,y,z$) are the strain and wavevector components, respectively. $\Delta_1$ is the crystal-field energy parameter, while $\Delta_2$ and $\Delta_3$ are spin-orbit energy parameters. The basis set used for this Hamiltonian spans six $p$-orbital states coupled with spin:
\begin{align}
  \phi_1^h &= \tfrac{1}{\sqrt{2}}\,\ket{X+iY,\uparrow}, \label{eq:b1}\\
  \phi_2^h &= \tfrac{1}{\sqrt{2}}\,\ket{X+iY,\downarrow}, \label{eq:b2}\\
  \phi_3^h &= \ket{Z,\uparrow}, \label{eq:b3}\\
  \phi_4^h &= \ket{Z,\downarrow}, \label{eq:b4}\\
  \phi_5^h &= \tfrac{1}{\sqrt{2}}\,\ket{X-iY,\uparrow}, \label{eq:b5}\\
  \phi_6^h &= \tfrac{1}{\sqrt{2}}\,\ket{X-iY,\downarrow}. \label{eq:b6}
\end{align}
These orbital-spin basis states allow the model to resolve polarization components. In particular, the $\ket{X\pm iY}$ states dominate TE-polarized transitions, while $\ket{Z}$ states contribute to TM polarization.

In table~\ref{tab:kp}, effective mass parameters were calculated based on the analytical expressions derived in Ref.~\cite{chuang1996}. In this work, the reported $A_1-A_6$ parameters of GaN and AlN were substituted into these expressions to obtain the direction-dependent effective masses for the CH$_1$ and CH$_3$ band. These effective masses were then used in the LL model to resolve the TE- and TM-polarized optical components.

\begin{table}[t]
\caption{\label{tab:kp}Parameters for the $6\times 6$ $k\cdot p$
matrix.}
\begin{ruledtabular}
\begin{tabular}{lcc}
  \textbf{Parameter} & \textbf{GaN} & \textbf{AlN} \\
  \hline
  $A_1$ & $-6.56^a$ & $-3.86^c$ \\
  $A_2$ & $-0.91^a$ & $-0.25^c$ \\
  $A_3$ & $5.65^a$  & $3.58^c$ \\
  $A_4$ & $-2.83^a$ & $-1.32^c$ \\
  $A_5$ & $-3.13^a$ & $-1.47^c$ \\
  $A_6$ & $-4.86^a$ & $-1.64^c$ \\
  $D_1$ (eV) & $-1.7^b$  & $-17.1^c$ \\
  $D_2$ (eV) & $6.30^b$  & $7.9^c$ \\
  $D_3$ (eV) & $8.0^b$   & $8.8^c$ \\
  $D_4$ (eV) & $-4.0^b$  & $-3.9^c$ \\
  $D_5$ (eV) & $-4.0^b$  & $-3.4^c$ \\
  $D_6$ (eV) & $-5.5^b$  & $-3.4^c$ \\
  $\Delta_1$ (eV) & $0.022^b$ & $-0.169^c$ \\
  $\Delta_2,\Delta_3$ (eV) & $0.005^b$ & $0.00633^c$ \\
  $a_c$ (eV) & $-4.50$ & $-4.50$ \\
  $m_{\ket{X\pm iY},x}^*/m_0$ & 0.267 & 0.637 \\
  $m_{\ket{X\pm iY},y}^*/m_0$ & 0.267 & 0.637 \\
  $m_{\ket{X\pm iY},z}^*/m_0$ & 1.0989 & 3.57 \\
  $m_{\ket{Z},x}^*/m_0$ & 1.0989 & 4.0 \\
  $m_{\ket{Z},y}^*/m_0$ & 1.0989 & 4.0 \\
  $m_{\ket{Z},z}^*/m_0$ & 0.1524 & 0.2590 \\
\end{tabular}
\end{ruledtabular}
\noindent$^a$Ref.~\citenum{chuang1996},
$^b$Ref.~\citenum{park2007},
$^c$Ref.~\citenum{vurgaftman2003}.
\end{table}

\subsection{\label{sec:ddcc}Drift-Diffusion Framework}

The DDCC formalism solves the standard Poisson equations and drift-diffusion equations in full 3D based on the FEM method, while self-consistently solving for the total charge density, electrostatic potential ($\psi$), and quasi-Fermi levels~\cite{li2016,yang2014}. This framework provides the spatial distributions of electrostatic potential, carrier densities, recombination rates, and internal quantum efficiency under steady-state operation:
\begin{equation}
  \label{eq:poisson}
  \nabla\cdot(\varepsilon\nabla\psi) = -q\!\left(p - n + N_D^+ - N_A^- \pm\rho_\text{pol}\right),
\end{equation}
where $\psi$ is the electrostatic potential. For conduction band potential $E_c$ and valence band $E_v$, they would be parallel to $-q\psi$. $n$ and $p$ are the electron and hole densities, $N_A^-$ and $N_D^+$ are the ionized acceptor and donor concentration, and $\rho_\text{pol}$ denotes the polarization charge density arising from divergence of the total polarization (sum of spontaneous and piezoelectric polarization). $\varepsilon$ is the dielectric constant.

\subsection{\label{sec:LL}Multi-Band Localization Landscape Model}

We incorporate the \LL{} (LL) model self-consistently into the coupled Poisson and drift-diffusion equations, enabling quantum confinement and
tunneling effects to be captured efficiently~\cite{filoche2017a,li2017,piccardo2017,arnold2016}. In the LL approach, quantum confinement effects in each relevant band are approximately obtained from a linear elliptic partial differential equation of the form
\begin{equation}
  \label{eq:LL}
  \hat{H}\,u = 1,
\end{equation}
where $\hat{H}$ is the single-particle Hamiltonian operator containing the effective mass and potential landscape for that band. The solution $u$ defines an
effective confinement potential $W = 1/u$ that governs localization, effective band edges, and densities of states.

For electrons, the LL equation is
\begin{equation}
  \label{eq:LLe}
  \left(-\frac{\hbar^2}{2m_e^*}\nabla^2 + E_c + \Delta E_{c,\text{strain}}\right)
  u_e = 1,
\end{equation}
where $m_e^*$ is the electron effective mass, $E_c$ is the conduction-band profile, and $\Delta E_{c,\text{strain}}$ is the strain-induced energy shift.

For the valence band, to include strain-induced valence-band deformation potentials shift for the $\ket{X\pm iY}$ and $\ket{Z}$ bands into the multi-band landscape solver, we applied the deformation term to the potential term $E_v$ for the two different bands. For simplicity, CH$_1$ and CH$_2$ are close in energy, and we only solve the CH$_1$ band and the CH$_3$ band separately to save computing costs.

These bands are treated within the LL framework as two independent scalar problems:
\begin{align}
  &\Bigg(-\frac{\hbar^2}{2m_{h,\text{CH1}}^*}\nabla^2 + E_v + \Delta_1 \nonumber\\
    &~~~+ \Delta_2 + \Delta E_{v,\text{strain,CH1}} \Bigg)
    u_{h,\text{CH1}} = 1, \label{eq:LLh1}\\
  &\left(-\frac{\hbar^2}{2m_{h,\text{CH3}}^*}\nabla^2 + E_v
    + \Delta E_{v,\text{strain,CH3}}\right)
    u_{h,\text{CH3}} = 1, \label{eq:LLh3}
\end{align}
where $m_{h,\text{CH}i}^*$ are the hole effective masses for CH$_1$ and CH$_3$. The effective masses in different direction are extracted from the \kp{} method to keep the comparison to \kp{} with the same parameters. $E_v$ is the valence band potential calculated by the Poisson equation. $\Delta E_{v,\text{strain,CH}i}$ are the strain-induced deformation potentials in CH$_1$--CH$_3$:
\begin{equation}
  \label{eq:dEc}
  \Delta E_{c,\text{strain}} =
  a_c\!\left(\varepsilon_{xx}+\varepsilon_{yy}+\varepsilon_{zz}\right)
  \times r_{E_c},
\end{equation}
\begin{multline}
  \label{eq:dEv1}
  \Delta E_{v,\text{strain,CH1}} =
  a_c\!\left(\varepsilon_{xx}+\varepsilon_{yy}+\varepsilon_{zz}\right)
  \times(r_{E_c}-1) \\
  +(D_1+D_3)\varepsilon_{zz}
  +(D_2+D_4)(\varepsilon_{xx}+\varepsilon_{yy}),
\end{multline}
\begin{multline}
  \label{eq:dEv3}
  \Delta E_{v,\text{strain,CH3}} =
  a_c\!\left(\varepsilon_{xx}+\varepsilon_{yy}+\varepsilon_{zz}\right)
  \times(r_{E_c}-1) \\
  + D_1\varepsilon_{zz} + D_2(\varepsilon_{xx}+\varepsilon_{yy}),
\end{multline}
where $D_1$--$D_4$ are the valence-band deformation potential coefficients in the \kp{} method, $a_c$ is the hydrostatic deformation potential. $r_{E_c} = 0.63$ is the conduction band offset ratio~\cite{filoche2017a,chang2017}. The strain-induced potential difference can be divided into two contributions: the strain-caused valence-band deformation potential from $D_1$--$D_6$ in \kp{}, and the strain-caused band gap change from the hydrostatic deformation potential.

Solving Eqs.~(\ref{eq:LLe})--(\ref{eq:LLh3}) on the 3D mesh yields the landscape functions $u_e$ and $u_{h,\text{CH}i}$, from which effective confinement potentials $W_e = 1/u_e$ and $W_{h,\text{CH}i} = 1/u_{h,\text{CH}i}$ are constructed. These effective potentials define spatially varying effective band edges and bandgaps used together with the 3D-DDCC carrier distributions to compute electroluminescence:
\begin{equation}
  \label{eq:Egeff}
  E_g^\text{eff} = \frac{1}{u_e} - \max\!\left(\frac{1}{u_{h,\text{CH}_i}}\right).
\end{equation}

By this substitution, the model directly captures the effects of quantum confinement and disorder.

The drift-diffusion model is described through the following set of coupled equations:
\begin{align}
  n =& ~\int_{1/u_e}^{+\infty} N_c(E)\,f_n(E)\,dE, \label{eq:n}\\
  p =& ~2\int_{-\infty}^{1/u_{h,\text{CH}_1}} N_{v,\text{CH}_1}(E)\,f_p(E)\,dE \nonumber\\
      &~+ \int_{-\infty}^{1/u_{h,\text{CH}_3}} N_{v,\text{CH}_3}(E)\,f_p(E)\,dE,
      \label{eq:p}\\
  \vec{J}_n =& ~n\mu_n\nabla E_{Fn}, \label{eq:Jn}\\
  \vec{J}_p =& ~p\mu_p\nabla E_{Fp}, \label{eq:Jp}\\
  \nabla\cdot\vec{J}_{n,p} =&~ -q(R-G), \label{eq:cont}\\
  R =& ~\frac{np-n_i^2}{\tau_{n0}(p+n_i)+\tau_{p0}(n+n_i)} \nonumber\\
      &~+ B_0 np + C_0(n^2p+np^2), \label{eq:R}
\end{align}
where $E_{Fn}$ and $E_{Fp}$ are the quasi-Fermi energies for electrons and holes, respectively. $f_n$ and $f_p$ are the Fermi distribution functions for electrons and holes, $\vec{J}_{n,p}$ are the electron and hole current densities, $\mu_{n,p}$ is the mobility, $R$ is the total recombination rate. The coefficients $\tau_{n0,p0}$ denote the Shockley--Read--Hall (SRH) lifetimes; $B_0$ and $C_0$ correspond to radiative and Auger recombination, respectively. G are the carrier generation rates. Hole densities are the summation of the CH$_1$ band (multiplied by a factor of two to include CH$_2$) and the CH$_3$ band. These equations are solved together with the continuity equations, including SRH, radiative, and Auger recombination, using the parameters listed in Table~\ref{tab:transport}.

\subsection{\label{sec:WW}Wigner--Weyl Emission Spectrum}

To further quantify the emission response, we adopt a \WW{} (WW) phase‑space formalism~\cite{banon2022}, which provides a semiclassical description of optical emission in spatially non‑uniform systems. In addition, the applicability of the LL framework to optical-transition calculations in superlattice structures has been  validated~\cite{Wu2026}. Within this framework, the emission rate takes a form analogous to that derived from Fermi’s golden rule, but is evaluated locally in phase space rather than between global eigenstates. A key distinction from the conventional formulation is that the effective bandgap is allowed to vary spatially. Specifically, the local bandgap is defined as the difference between the electron effective potential and the hole effective quantum potential as shown below:
\begin{equation}
  \label{eq:Egeff1}
  E_{{g,\text{CH}_1}}^\text{eff} = \frac{1}{u_e(r)} - \left(\frac{1}{u_{h,\text{CH}_1}(r)}\right)\\
\end{equation}
\begin{equation}
  \label{eq:Egeff3}
  E_{{g,\text{CH}_3}}^\text{eff} = \frac{1}{u_e(r)} - \left(\frac{1}{u_{h,\text{CH}_3}(r)}\right).
\end{equation}

Which naturally accounts for band bending and quantum confinement effects. This position‑dependent bandgap enables a local evaluation of the emission rate in systems where strong spatial variations of the band structure are present. Then the emission rate can be represent as
\begin{widetext}
\begin{align}
  W_\text{WWL}^\text{TE}(\hbar\omega) &=
  \frac{2}{V}\int_\Omega
  \frac{1}{3}\frac{q^2 n_r \hbar\omega}{\pi m_0 \varepsilon_0 c^3 \hbar^2}
  \frac{E_p}{2} \cdot \frac{3}{2}
  \left(\frac{2m_{r,\text{CH}_1}(\vec{r})}{\hbar^2}\right)^{3/2}
  \!\!\left(\hbar\omega - E_{g,\text{CH}_1}^\text{eff}(\vec{r})\right)_+^{1/2}
  f_e f_h\,d^3r, \label{eq:TE}\\
  W_\text{WWL}^\text{TM}(\hbar\omega) &=
  \frac{1}{V}\int_\Omega
  \frac{1}{3}\frac{q^2 n_r \hbar\omega}{\pi m_0 \varepsilon_0 c^3 \hbar^2}
  \frac{E_p}{2} \cdot \frac{3}{2}
  \left(\frac{2m_{r,\text{CH}_3}(\vec{r})}{\hbar^2}\right)^{3/2}
  \!\!\left(\hbar\omega - E_{g,\text{CH}_3}^\text{eff}(\vec{r})\right)_+^{1/2}
  f_e f_h\,d^3r, \label{eq:TM}
\end{align}
\end{widetext}
where $E_p$ is the Kane energy associated with the momentum matrix element, and $n_r$ is the refractive index of the material. $m_0$ is the electron rest mass, $c$ is the velocity of light, and $m_{r,\text{CH}_1}$ and $m_{r,\text{CH}_3}$ are the local reduced effective masses. $f_e$ and $f_h$ denote the electron and hole occupation factors, respectively. The EL emission from the $CH_3$ and $CH_1$ bands can be calculated separately to obtain different TM/TE ratios. These integrals allow us to efficiently compute electroluminescence (EL) characteristics while retaining sensitivity to nano-scale compositional disorder.

To compare the simplified LL model with the \kp{} result, we solve the 3D \kp{} model after the program converges. After obtaining the band structure, we calculate the eigenfunctions and eigenstates of electrons and holes from the \kp{} calculations. The spontaneous emission rate, which determines the EL intensity, is evaluated using
\begin{widetext}
\begin{equation}
  \label{eq:EL_kp}
  R = \int
  \frac{q^2 n_r \hbar\omega}{\pi m_0^2 \varepsilon_0 c^3 \hbar^2}
  \cdot\frac{1}{3}
  \sum_{i,j}\left|\braket{\phi_i u_s}{\hat{a}\cdot\hat{p}}{\phi_j u_p}\right|^2
  \times f^e(E_i^e)\,f^h(E_j^h)
  \cdot\frac{1}{\sqrt{2\pi}\,\sigma}
  \exp\!\left[-\frac{(E_{ij}-\hbar\omega)^2}{2\sigma^2}\right]
  d(\hbar\omega),
\end{equation}
\end{widetext}
where R is the Electroluminescence (EL) intensity generated by radiative recombination; $i$ and $j$ label electron and hole states, respectively; $1/3$ accounts for the EL intensity in one propagation direction; $\braket{\phi_i u_s}{\hat{a}\cdot\hat{p}}{\phi_j u_p}$ describes different state transitions and their wavefunction overlap under perturbation theory; $\phi_i$ and $\phi_j$ are the $i$th and $j$th envelope wavefunctions of the conduction and valence bands; $u_s$ and $u_p$ are the core states of $s$ and $p$ orbitals; $\hat{a}$ is the electric field polarization direction; $\hat{p}$ is the momentum operator; $E_{ij}$ is the transition energy; $\sigma$ is the standard deviation of the Gaussian broadening function; $f^e(E_i^e)$ and $f^h(E_j^h)$ are Fermi--Dirac distributions for electrons and holes in different eigenstates.

The TE/TM polarization of the emitted light is determined by the valence-band character of the hole channels involved in radiative recombination. From the \kp{} eigenvectors $\phi_{\text{CH}i}$, we extract the fractions associated with the $\ket{X\pm iY}$ (TE-like) and $\ket{Z}$ (TM-like) basis states for each channel. These fractions are used as weighting factors to decompose the total EL intensity into TE- and TM-polarized components, $I_\text{TE}$ and $I_\text{TM}$, by summing over all channels and spatial contributions. Finally, the polarization ratio is defined as
\begin{equation}
  \label{eq:PR}
  \text{PR} = \frac{I_\text{TM} - I_\text{TE}}{I_\text{TM} + I_\text{TE}},
\end{equation}
providing a quantitative measure of emission anisotropy as a function of wavelength.

% ==================================================================
\section{\label{sec:results}Results and Discussion}
% ==================================================================

\begin{table*}[t]
\caption{\label{tab:transport}Parameters adopted for each region in this study. All values remain constant throughout the simulations.}
\begin{ruledtabular}
\begin{tabular}{lcccccc}
  \textbf{Region} & \textbf{Thickness} (nm)
    & \textbf{Doping} ($10^{18}$~cm$^{-3}$)
    & $e^-/h^+$ \textbf{mobility} (cm$^2$/Vs)
    & $B_0$ ($10^{-11}$~cm$^3$/s)
    & $C_0$ ($10^{-30}$~cm$^6$/s)
    & $\tau_{n,p}$ (ns) \\
  \hline
  p-GaN         & 33 & 50  & 5/2   & 2 & 0.2 & 10 \\
  p-\AlGaN      & 12 & 50  & 5/2   & 2 & 0.2 & 10 \\
  p-AlGaN       & 10 & 50  & 5/2   & 2 & 0.2 & 10 \\
  LQB           & 10 & 5   & 5/2   & 2 & 0.2 & 10 \\
  QB            & 10 & 0.1 & 100/5 & 2 & 0.2 & 10 \\
  QW            &  2 & 0.1 & 100/5 & 2 & 0.2 & 10 \\
  n-first QB    & 10 & 1  & 21.9/2& 2 & 0.2 & 10 \\
  n-AlGaN       & 40 & 50  & 21.9/2& 2 & 0.2 & 10 \\
\end{tabular}
\end{ruledtabular}
\end{table*}

\begin{figure}[t]
  \includegraphics[width=0.9\columnwidth]{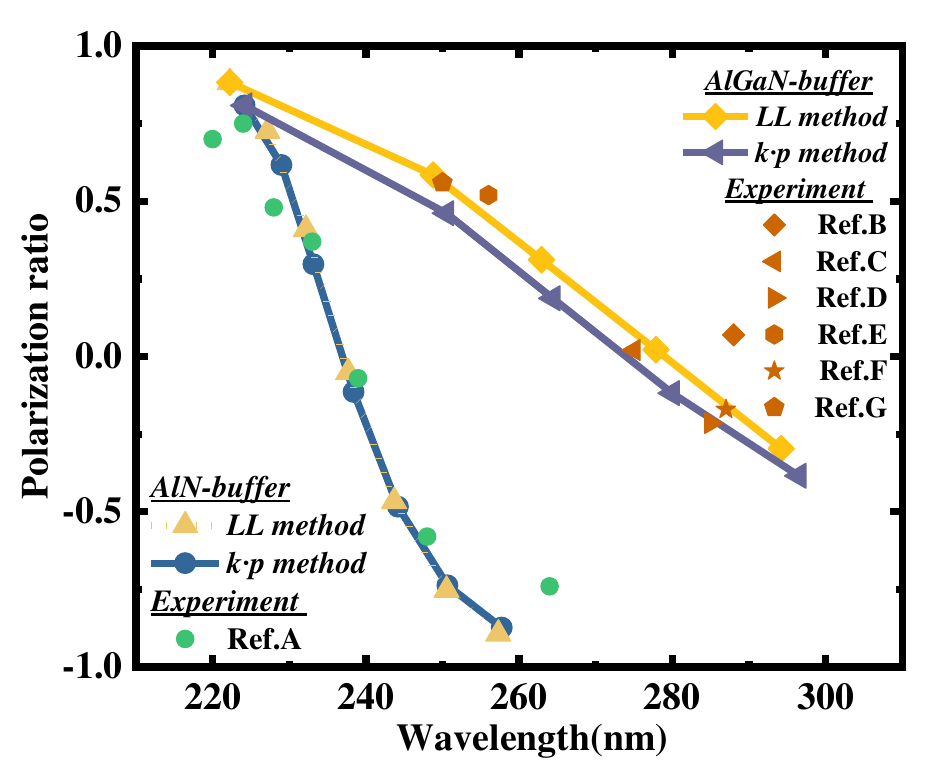}
  \caption{\label{fig:PR}Wavelength-dependent polarization ratio (220--300~nm) of AlGaN UVC LEDs for AlGaN- and AlN-buffer cases. The   LL (\WW{}) model matches the $6\times 6$ \kp{} trend and reproduces the buffer-dependent TE-TM crossover, consistent with literature experiments. Among the experimental references, Ref. A (green point) is for the AlN-buffer case, whereas Refs. B–G (orange-brown point) are for the AlGaN-buffer case. (Ref.~A~\cite{guttmann2019}, Ref.~B~\cite{kolbe2010}, Ref.~C~\cite{lee2016}, Ref.~D~\cite{wang2016}, Ref.~E~\cite{gong2024}, Ref.~F~\cite{wang2019}, Ref.~G~\cite{tian2024})}
\end{figure}

Figure~\ref{fig:PR} benchmarks the polarization ratio from the localization landscape (LL) model, combined with the \WW{} formalism, against a full $6\times 6$ \kp{} matrix solver for AlGaN UVC LED structures spanning the 220--300~nm emission range (evaluated under a representative operating condition). The optical polarization ratio exhibits a clear wavelength-dependent evolution and a buffer-sensitive TE-TM crossover: devices with an AlN buffer show relatively weaker TM polarization, whereas the AlGaN-buffer case maintains a higher TM polarization ratio over a broader wavelength range before gradually approaching the crossover at longer wavelengths. This behavior is primarily attributed to the stronger quantum-well compressive stress induced by the AlN buffer, which yields a larger strain-driven energy shift. As illustrated in Fig.~\ref{fig:banddg}(b), the $\ket{Z}$ band exhibits a larger energy downshift under compressive strain, modifying the valence-band ordering and reducing the CH$_3$-related TM transition strength.

Importantly, the LL curves track the \kp{} results closely for both buffer configurations, indicating that LL can correctly desribe the multiband band-mixing trends that govern polarization selection rules in these alloy-disordered heterostructures. In addition, both simulation approaches show good overall agreement with the reported experimental data, confirming that the present model can correctly reproduce the wavelength-dependent polarization behavior observed in AlGaN UVC LEDs. The close agreement between LL and \kp{}, together with the consistency with reported experimental points, supports the use of LL computationally efficient at optical light polarization-related predictions. 

\begin{figure}[t]
  \includegraphics[width=0.9\columnwidth]{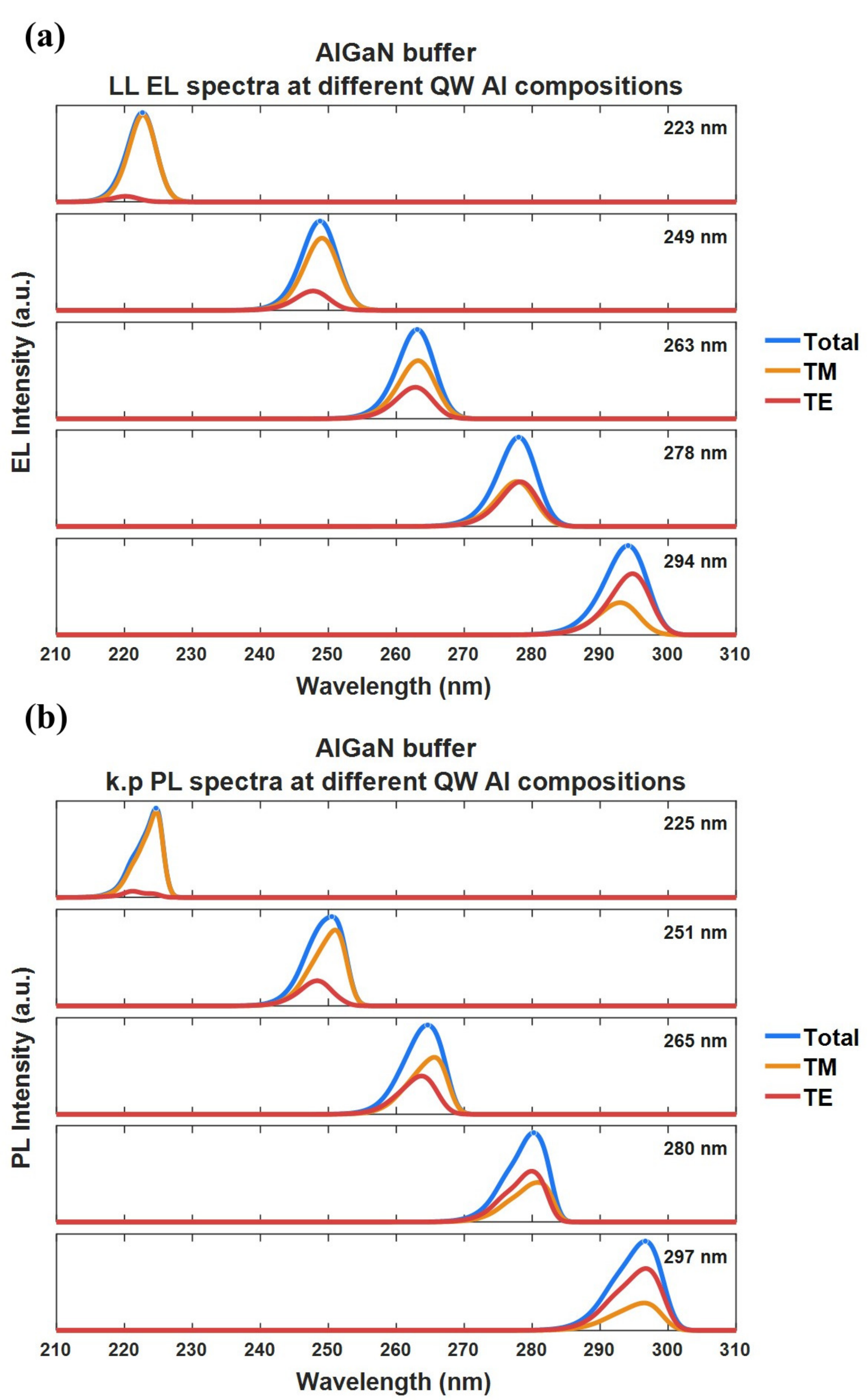}
  \caption{\label{fig:spectra}TE- and TM-resolved optical spectra calculated using the (a) LL and (b) \kp{} methods for the AlGaN-buffer structure.}
\end{figure}

To provide a more intuitive picture of the polarization-ratio evolution, Fig.~\ref{fig:spectra} presents the TE- and TM-resolved optical spectra calculated using the LL and $6\times 6$ \kp{} methods for the AlGaN-buffer structure.  Both methods show the same wavelength-dependent change in the relative TE and TM emission components. With increasing QW Al composition, the Z band gradually becomes the dominant hole ground state, enhancing the TM-polarized optical transition. As a result, the TM component becomes increasingly pronounced as the emission wavelength shifts toward shorter wavelengths. This spectral change visually explains the wavelength-dependent polarization-ratio trend discussed above.

Despite its reduced numerical complexity, the LL approach retains the essential disorder-induced localization physics and captures the dominant operating-condition effects (e.g., carrier-induced band filling and field screening) that shape the polarization behavior. The LL-based approach reduces the simulation time from to thousands of hours to a few hours while maintaining high accuracy in predicting emission spectra and carrier transport, as shown in Table~\ref{tab:cost}.As a result, the LL model provides a practical pathway for fast design-space exploration and buffer/structure optimization in UVC LEDs where full \kp{} simulations would be prohibitively expensive.

\begin{table*}[t]
\caption{\label{tab:cost}Estimated total computational time for 3D LL vs. 3D $k\cdot p$.}
\begin{ruledtabular}
\begin{tabular}{lccc}
  \textbf{Solver} & \textbf{Per-iteration time}  & \textbf{Mesh Nodes number} & \shortstack{\textbf{Under the same } \\ \textbf{number of mesh nodes}} \\
  \hline
  Poisson + DD ($n$, $p$) + 3 LL equations  & $\sim$32~min 34~s & 2\,340\,509 & $\sim$0.5~h \\
  Only solve $k\cdot p$ method(1200 eigenstate)
  & $\sim$200~h for 1~QW  & 208\,860 (1 QW)   & $\sim$2240~h \\
\end{tabular}
\end{ruledtabular}
\end{table*}

Table~\ref{tab:cost} compares the estimated computational costs between the 3D LL and 3D \kp{} methods. Note that the matrix size of $6\times6$ \kp{} method is too large if we want to compute the whole device domain. If the mesh node number is N. The matrix size of Hamiltonian for \kp{}  would be $(6N)^2$ , which is much larger than the original $N^2$  problem. Of course, the matrix is sparse matrix where we didn’t really need such a large matrix. However, computing resources required for \kp{} is much larger than the one band Scrodinger equations. Hence, the eigenvalues and eigenfunctions of each QW are solved separately to save computing memory. 

In particular, for the \kp{} calculation listed in Table~\ref{tab:cost}, 1200 hole eigenstates were explicitly solved for constructing the EL spectrum. Each hole eigenstate contains six coupled valence-band components over the 3D mesh, and the corresponding eigenfunctions must be obtained to calculate the electron–hole wavefunction overlap and TE/TM optical transition strengths.

Notably, despite utilizing a much larger number of nodes (2.34 million vs. 0.21 million), the Poisson and drift-diffusion solver + LL method completes the calculation in significantly less time (~0.5 hour) compared to the \kp{} method (~2240 hours). Note that if we want to solve \kp{} for whole active region at the same time, it might need even longer times. These computing are calculated based on 8 cores x86 based CPU. The PARDISO libaray in Intel MKL and ARPACK solver is used for solving the sparse matrix.

\begin{figure}[t]
  \includegraphics[width=\columnwidth]{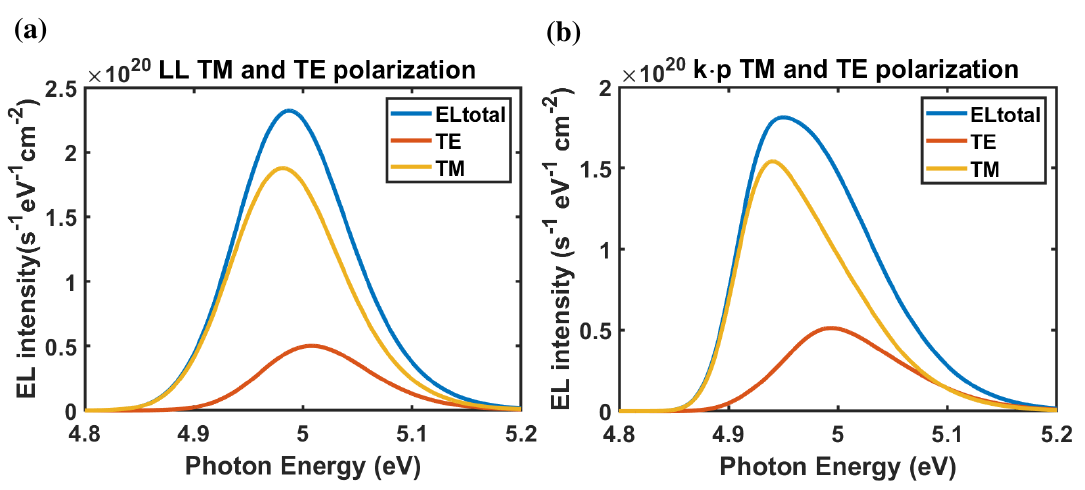}
  \caption{\label{fig:EL}The TE/TM-resolved EL spectra at 249~nm (7.2~A$\cdot$cm$^{-2}$) for the AlGaN-buffer structure with an Al$_{0.6}$Ga$_{0.4}$N QW, comparing (a) LL-WW and (b) $k\cdot p$ calculations; TE, TM, and total emission components are included.}
\end{figure}

The electroluminescence (EL) spectrum derived by the LL model accurately reproduces both the peak energy and the spectral shape obtained from the \kp{} calculation, demonstrating strong consistency between the two approaches, as shown in Fig.~\ref{fig:EL}.

Figure~\ref{fig:EL} shows the computed EL spectrum at 249~nm for the AlGaN substrate, where the strengths of TM and TE waves are at the crossover point. In the \kp{} calculation, we solved for 1200 hole eigenstates to construct the EL spectrum; however, these 1200 eigenstates span an energy range of only about 0.11 eV. In addition, The \kp{} results solve limited eigenstates in a limited space. Hence, some step-like peaks appear, whereas more random seeding numbers are needed to realize good statistics. These results confirm that the LL approach preserves essential valence-band-mixing physics while reducing the computational cost by more than an order of magnitude.

\begin{figure}[t]
  \includegraphics[width=0.8\columnwidth]{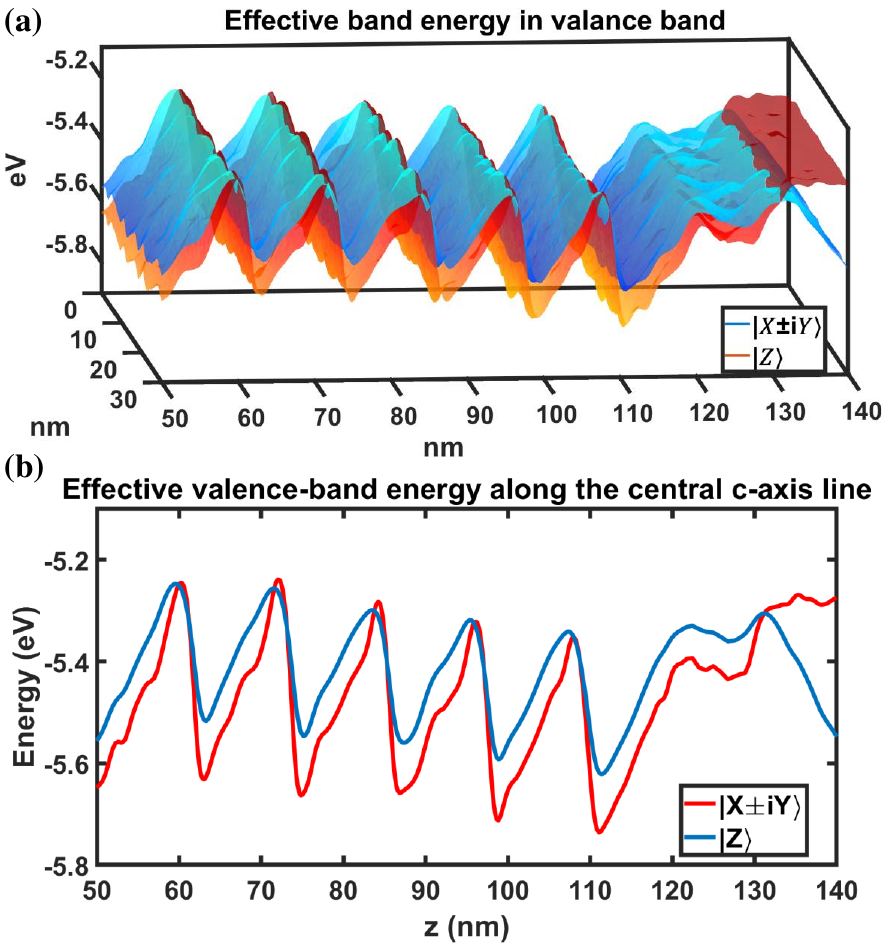}
  \caption{\label{fig:landscape}(a) Three-dimensional effective valence-band energy profiles obtained by the present method, revealing the mixing between the $\ket{X\pm iY}$ and $\ket{Z}$ valence-band components. (b)~Line-cuts along the growth direction comparing QW stacks with AlN buffers, indicating that the valence-band character varies spatially due to strain and alloy  fluctuations. This case at 238~nm for the AlN-buffer structure with an Al$_{0.65}$Ga$_{0.35}$N QW, we apply forward bias  5.3V and current density is 12.7~A$\cdot$cm$^{-2}$.}
\end{figure}

Figure~\ref{fig:landscape} uses the LL framework to quantify how strain reshapes the effective valence-band energy landscape in quantum wells(QWs) through the multiband deformation-potential indicator $1/u_v$. It exhibits spatial changes due to the alloy fluctuation in both the QB and QW. This case is plotted for the 65\%~Al QW under an AlN buffer layer. Panel (a) is the 2D plot of potential along the $x$--$z$ plane, and (b) is the 1D plot along the $z$-direction at x and y at the QW center. Before strain is applied, the dominant band is the $\ket{Z}$ state. However, after compressive strain, the $\ket{XY}$ band becomes sometimes locally the ground state. Fig.~\ref{fig:landscape}(b) shows the 1D plot in the middle of the simulation domain in the $z$-direction (c-axis). With an AlN buffer, the $1/u_v$ profiles are shifted overall toward lower energy (more negative values), indicating that the buffer material alters the residual stress state across the QW/QB stack, thereby changing both the strength of the spatial valence-band-edge modulation and the energy separation between the $\ket{X\pm iY}$ and $\ket{Z}$ components. This qualitative behavior is consistent with the schematic trend illustrated in Fig.~\ref{fig:banddg}(b): compressive in-plane stress leads to a downward shift of valence-band edges, especially for the $\ket{Z}$ band. The QW suffers more effect of compressive strain than QB so that effective QW is shallower for holes. Consequently, the buffer-dependent differences in $1/u_v$ are expected to directly influence optical selection rules, and thus impact the TE/TM polarization tendency, providing a practical physical basis for polarization control and structural optimization.

To quantify how compressive strain modifies hole confinement and injection in AlGaN-based UVC LEDs, we disable the strain-induced valence-band deformation term in the multiband LL framework and compare the results against the strained cases. In wurtzite AlGaN MQWs, compressive strain changes the relative energies of the dominant valence subspace, effectively reducing the valence-band offset $\Delta E_v$ experienced by holes inside the QWs. As a result, the hole QW becomes shallower, and the barrier limiting hole transport across the QW/QB stack is reduced. In addition, when the $\ket{Z}$-like band becomes more relevant for hole transport, its smaller effective mass along the growth direction further facilitates vertical hole conduction through the MQWs. The hole is more easier tunneling through the barriers in the LL model with a lighter effective mass in the z-directions. 

\begin{figure}[t]
  \includegraphics[width=0.8\columnwidth]{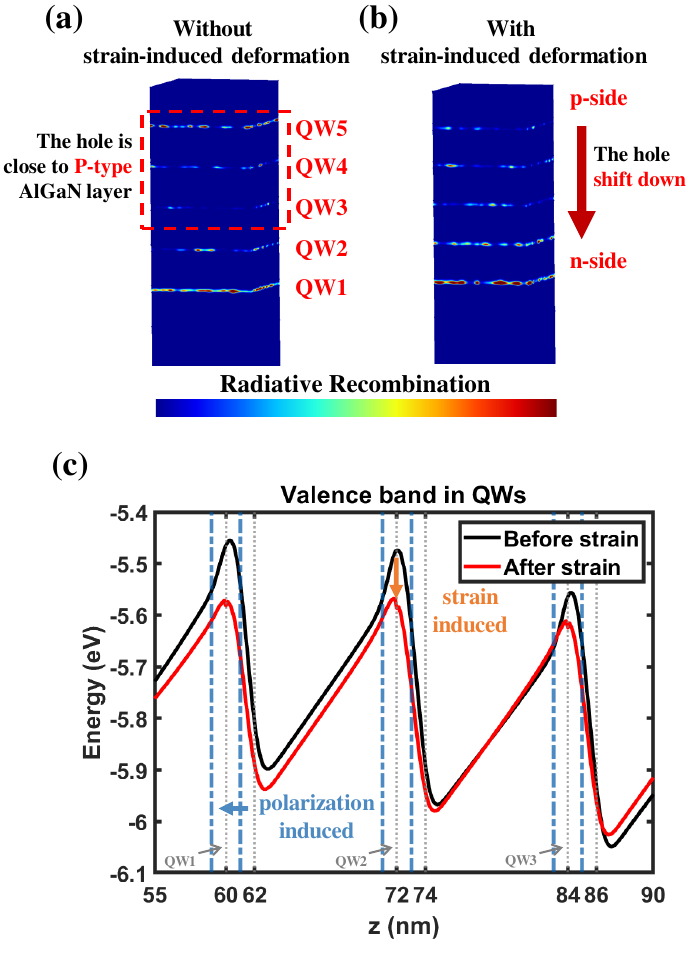}
  \caption{\label{fig:recomb}Radiative recombination maps at 223~nm and $J \approx 100$~A/cm$^2$ (a) without and (b) with the strain-induced valence-band deformation term, showing a downward shift of the radiative zone toward the n-side. (c) Valence-band profiles in the MQWs indicate reduced hole confinement (shallower QWs) under compressive strain. Figure~\ref{fig:banddg}(c) provides a schematic interpretation of this valence-band modification: polarization-induced band bending shifts the hole quantum well toward the QW/QB interface (blue arrow), while the strain-induced deformation term further reduces the hole well depth (orange arrow).}
\end{figure}

Figure~\ref{fig:recomb} directly shows the impact of compressive strain on the recombination profile. When the deformation term is not considered, holes remain preferentially accumulated near the p-side MQW region, resulting in a recombination zone closer to the p-type layers. After including compressive strain, holes move further and becomes less confined near the p-side, consistent with reduced hole confinement in the QWs (shallower hole QWs). Consequently, the overall radiative recombination region moves toward the n-type side, indicating more efficient hole penetration through the MQW stack.

To further interpret this behavior, Fig.~\ref{fig:banddg}(c) serves as a schematic illustration of the valence-band modification shown in Fig.~\ref{fig:recomb}(c). Owing to polarization-induced band bending, the hole accumulation position is shifted from the QW region toward the QW/QB interface, as indicated by the blue arrow. In addition, the simulations confirm that when stress is included the hole quantum well becomes shallower, as marked by the orange arrow, explaining the weakened hole confinement and the downward shift of the recombination zone.

\begin{figure}[t]
  \includegraphics[width=\columnwidth]{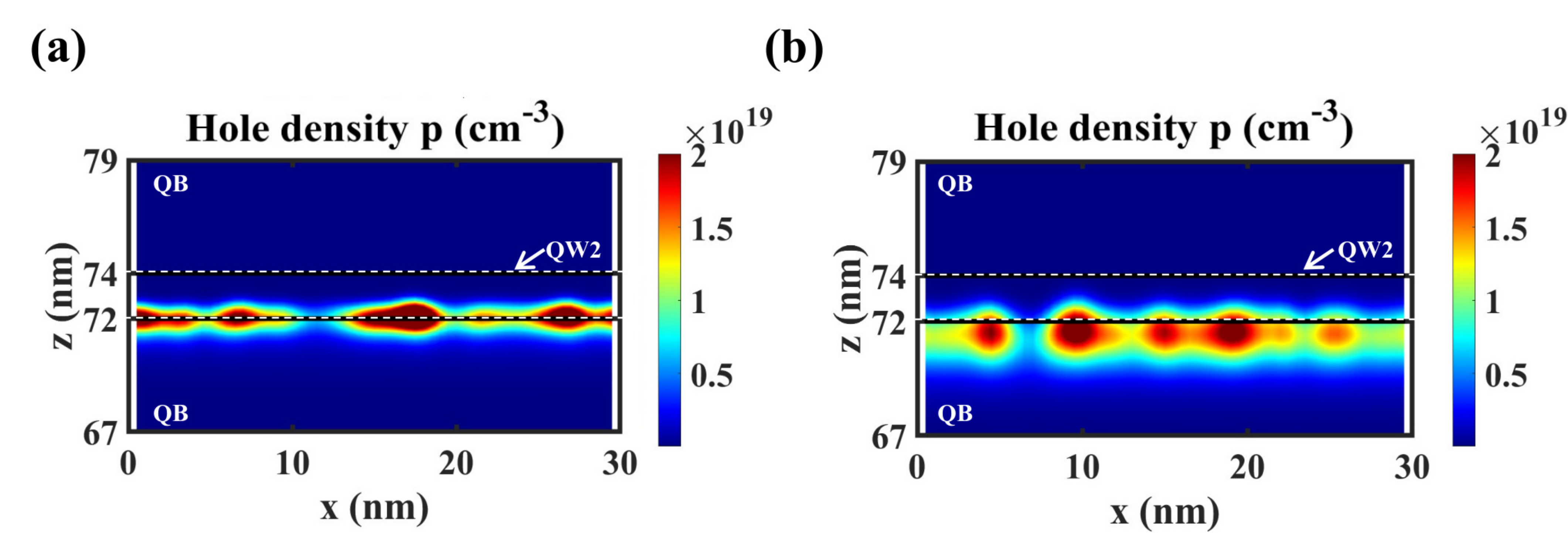}
  \caption{\label{fig:holes}Hole density at 223~nm and   $J \approx 100$~A/cm$^2$: (a) without and (b) with the deformation term. The hole fraction in barriers increases from $\sim$48\% to $\sim$75\%, indicating improved vertical hole transport through the MQWs.The region between the two dashed lines denotes QW2.}
\end{figure}

This trend is further supported by the hole density maps in Fig.~\ref{fig:holes}, where the strain-enabled case exhibits a substantially increased hole population inside the quantum barriers. To make this comparison quantitative, we evaluate the fraction of holes residing in the barrier region, $\eta_{QB} = p_{QB}/(p_{QB}+p_{QW})\times 100\%$, and find that $\eta_{QB}$ increases from approximately 48\% (without the deformation term) to approximately 75\% (with the deformation term). Because the polarization-induced band bending keep the hole accumulation position away from the center of the QW and toward the QW/QB interface, a significant portion of holes is consequently observed inside the QB region. The higher barrier population is a direct signature of reduced hole confinement and improved vertical transport, which promotes a more uniform hole supply across multiple wells.

\begin{figure}[t]
  \includegraphics[width=0.8\columnwidth]{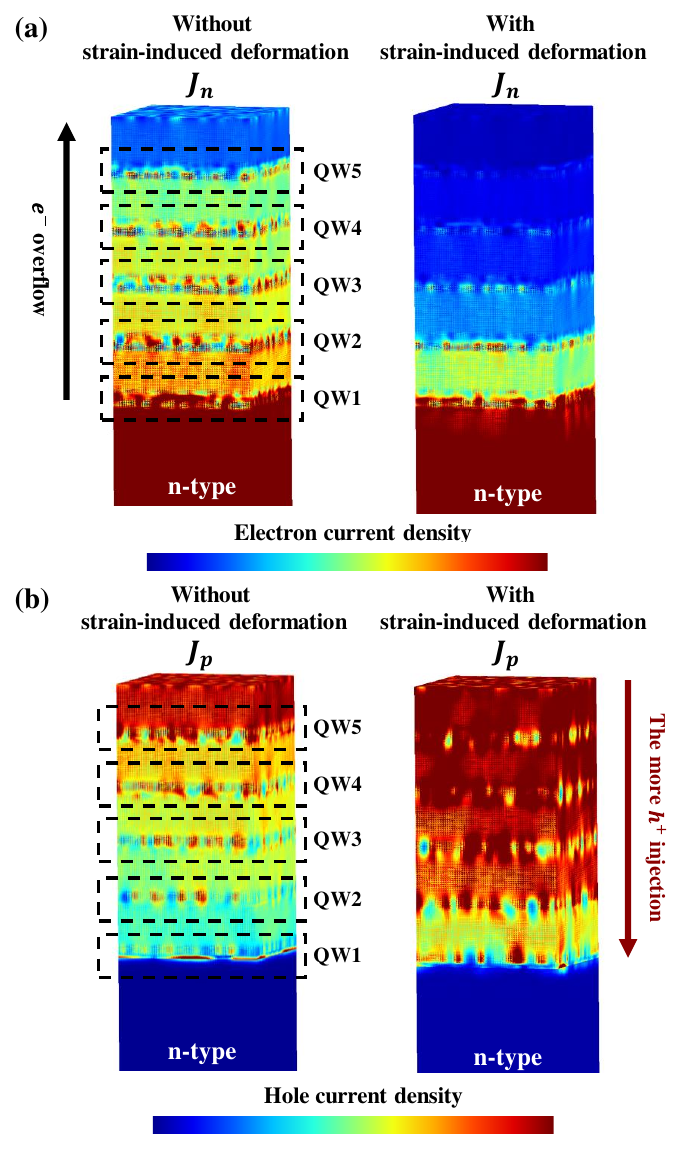}
  \caption{\label{fig:JJ}Electron and hole current-density maps at 223~nm and $J \approx 100$~A/cm$^2$, comparing cases without and with the deformation term. (a) electron current density (b) hole current density. Including strain enhances hole injection across QW1--QW5 and reduces electron leakage beyond the MQWs.}
\end{figure}

Finally, Fig.~\ref{fig:JJ} links the improved hole injection to electron leakage suppression. With strain-induced deformation enabled, the hole current density $J_p$ shows stronger capability to penetrate through the quantum barriers in the MQW region, which enhances electron-hole charge balance and increases the probability of radiative recombination within the active region. As a consequence, the electron current density $J_n$ exhibits reduced penetration beyond the MQWs, demonstrating suppressed electron overflow at high injection. These results confirm that properly accounting for compressive-strain-induced valence-band deformation is essential for accurately capturing hole transport physics and the resulting efficiency behavior in AlGaN UVC LEDs.

\begin{figure}[t]
  \includegraphics[width=\columnwidth]{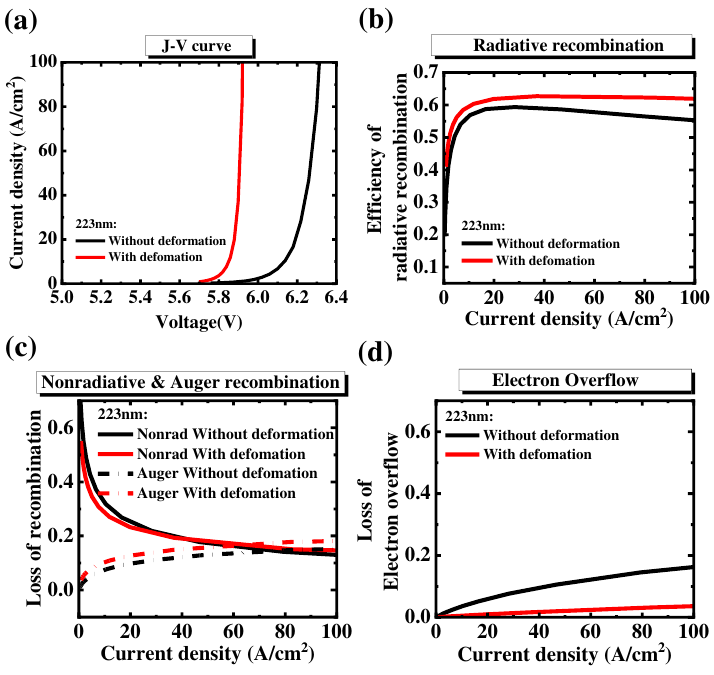}
  \caption{\label{fig:elec}Electrical and optical characteristics of UVC LEDs with and without the deformation term at 223~nm. (a) $J$--$V$ curves; (b) radiative recombination efficiency;  (c) nonradiative and Auger recombination loss; (d)~electron overflow.}
\end{figure}

Figure~\ref{fig:elec} shows the electrical and optical characteristics for the AlN quantum barrier (QB) at 223~nm,  $\text{Al}_x\text{Ga}_{1-x}\text{N}$ is considered with $x = 1.0$, resulting in a QB without alloy fluctuation. From the electrical characteristics obtained with the multi-band LL model, we can clearly observe decreased electron overflow decrease and IQE increase when the deformation term is included.

% ==================================================================
\section{\label{sec:conclusion}Conclusion}
% ==================================================================

In this work, we developed an efficient and physically rigorous simulation framework for AlGaN-based UVC LEDs by integrating the \LL{} (LL) model with the \WW{} formalism and a full 3D drift-diffusion solver. This hybrid approach captures the essential quantum effects---carrier localization, tunneling, band mixing, and disorder-induced potential fluctuations---while avoiding the high computational cost of solving the full 3D \kp{} eigenvalue problem. By incorporating strain-induced deformation-potential shifts into a multi-band LL formulation, the model accurately reproduces how compressive strain modifies the valence-band structure, reduces the effective hole-confinement barrier, and enhances vertical hole transport through the MQW stack. The improved hole injection helps suppress electron overflow and yields a more balanced carrier distribution, which is essential for high-efficiency UVC emission, especially at wavelengths below 250~nm.

The LL-based method also successfully predicts the wavelength-dependent TE/TM polarization behavior and matches the trends obtained from $6\times 6$ \kp{} calculations as well as experimental data. Despite the significantly reduced numerical complexity, the LL--WW approach preserves the key physical mechanisms governing polarization switching, alloy-induced localization, and strain-dependent band reordering.

Overall, this study demonstrates that the LL framework provides more than an order-of-magnitude reduction in computation time while maintaining high accuracy in predicting electrical, optical, and polarization characteristics of AlGaN UVC LEDs. This enables rapid design-space exploration and offers a practical tool for optimizing deep-UV emitters where strain, alloy disorder, and multiband interactions critically determine device performance.

% ==================================================================
\begin{acknowledgments}
This work was supported by the National Science and Technology Council (NSTC), Taiwan, under Grant Nos.\ 113-2124-M-002-013-MY3, 115-2221-E-002-103-MY2, and 112-2221-E-002-215-MY3. Prof.\ Yuh-Renn Wu was also supported by the LEAP Fellowship of the Foundation for the Advancement of Outstanding Scholarship Prof.\ M. Filoche was supported by a grant from the Simons Foundation (Grant No.\ 1027116). Support at the University of California, Santa Barbara (UCSB), was provided by the Solid State Lighting and Energy Electronics Center (SSLEEC), the Simons Foundation (Grant Nos.\ 601952 and 1027114 for Prof.\ J.S. Speck and C. Weisbuch, respectively), and the U.S. Air Force Office of Scientific Research (AFOSR) under Award No.\ FA2386-24-1-4050.
\end{acknowledgments}

% ==================================================================
\nocite{*}
\bibliography{UVCLED}

\clearpage
\onecolumngrid

\appendix
\section{Al composition tables}

\renewcommand{\thetable}{A\arabic{table}}
\setcounter{table}{0}

The detailed Al composition parameters for the simulated AlGaN-buffer and AlN-buffer UVC-LEDs are listed in Tables~\ref{tab:AlGaN_buffer} and~\ref{tab:AlN_buffer}, respectively.

\begin{table}[h!]
\caption{\label{tab:AlGaN_buffer}Al composition parameters of the AlGaN-buffer devices at different emission wavelengths.}
\begin{ruledtabular}
\begin{tabular}{lccccc}
  \textbf{Al composition (\%)} & \textbf{294~nm} & \textbf{278~nm} & \textbf{263~nm} & \textbf{249~nm} & \textbf{223~nm} \\
  \hline
  p-AlGaN                         & 0     & 0    & 0     & 0    & 0     \\
  p-$\mathrm{Al}_x\mathrm{Ga}_{1-x}\mathrm{N}$ & 37.81--0 & 45--0 & 52.5--0 & 70--0 & 87.5--0 \\
  p-AlGaN                         & 46.41 & 52.5 & 61.25 & 75   & 93.75 \\
  LQB                             & 50    & 60   & 70    & 80   & 100   \\
  QB                              & 50    & 60   & 70    & 80   & 100    \\
  QW                              & 30    & 40   & 50    & 60   & 80   \\
  n-first QB                      & 50    & 60   & 70    & 80   & 100   \\
  n-AlGaN                         & 50    & 60   & 70    & 80   & 100   \\
  n-contact                       & 50    & 60   & 70    & 80   & 100   \\
  buffer                          & 50    & 60   & 70    & 80   & 100   \\
\end{tabular}
\end{ruledtabular}

\end{table}

\begin{table}[h!]
\caption{\label{tab:AlN_buffer}Al composition parameters of the AlN-buffer devices at different emission wavelengths.}
\begin{ruledtabular}
\begin{tabular}{lcccccc}
  \textbf{Al composition (\%)} & \textbf{257~nm} & \textbf{251~nm} & \textbf{244~nm} & \textbf{238~nm} & \textbf{232~nm} & \textbf{228~nm} \\
  \hline
  p-AlGaN                         & 0     & 0     & 0    & 0     & 0     & 0     \\
  p-$\mathrm{Al}_x\mathrm{Ga}_{1-x}\mathrm{N}$ & 52.5--0 & 65.63--0 & 70--0 & 74.38--0 & 78.75--0 & 83.13--0 \\
  p-AlGaN                         & 61.25 & 70.31 & 75   & 79.69 & 84.38 & 89.06 \\
  LQB                             & 70    & 75    & 80   & 85    & 90    & 95    \\
  QB                              & 70    & 75    & 80   & 85    & 90    & 95    \\
  QW                              & 50    & 55    & 60   & 65    & 70    & 75    \\
  n-first QB                      & 70    & 75    & 80   & 85    & 90    & 95    \\
  n-AlGaN                         & 70    & 75    & 80   & 85    & 90    & 95    \\
  n-contact                       & 70    & 75    & 80   & 85    & 90    & 95    \\
  buffer                          & 100   & 100   & 100  & 100   & 100   & 100   \\
\end{tabular}
\end{ruledtabular}
\end{table}

\end{document}